\documentclass[3p,numbers,sort&compress]{elsarticle}
\usepackage{graphics}
\usepackage{amssymb}
\usepackage{url}
\usepackage{epsfig}
\usepackage{subfigure}
\usepackage{amsmath}
\usepackage{pifont}
\usepackage{natbib}
\usepackage{geometry}
\usepackage{fleqn}
\usepackage{pstricks}
\usepackage{color}
\newcommand{\dat}{X}
\begin{document}
\title{Characterizing Multi-Scale Self-Similar Behavior and Non-Statistical Properties of Fluctuations in Financial Time Series.}
\author[sg]{Sayantan Ghosh\corref{cor1}}
\ead{210556397@ukzn.ac.za}
\author[pm]{P. Manimaran}
\ead{rpmanimaran@gmail.com}
\author[pkp]{Prasanta K. Panigrahi}
\ead{pprasanta@iiserkol.ac.in}
\address[sg]{School of Physics, University of KwaZulu-Natal, Private Bag X54001, Durban 4000, South Africa.}
\address[pm]{Center for Mathematical Sciences, C R Rao Advanced Institute of Mathematics, Statistics and Computer Science, Uni. of Hyderabad Campus, Gachi Bowli, Hyderabad 500 046, India.}
\address[pkp]{Indian Institute of Science Education and Research-Kolkata, BCKV Campus Main Office, Mohanpur, West Bengal 741 252, India.}
\cortext[cor1]{Corresponding author}
\begin{abstract}
We make use of wavelet transform to study the multi-scale, self similar behavior and deviations thereof, in the stock prices of large companies, belonging to different economic sectors. The stock market returns exhibit multi-fractal characteristics, with some of the companies showing deviations at small and large scales. The fact that, the wavelets belonging to the Daubechies' (Db) basis enables one to isolate local polynomial trends of different degrees, plays the key role in isolating fluctuations at different scales. One of the primary motivations of this work is to study the emergence of the $k^{-3}$ behavior \cite{hes5} of the fluctuations starting with high frequency fluctuations. We make use of Db4 and Db6 basis sets to respectively isolate local linear and quadratic trends at different scales  in order to study the statistical characteristics of these financial time series. The fluctuations reveal fat tail non-Gaussian behavior, unstable periodic modulations, at finer scales, from which the characteristic $k^{-3}$ power law behavior emerges at sufficiently large scales. We further identify stable periodic behavior through the continuous Morlet wavelet.
\end{abstract}
\maketitle

\section{Introduction}
Study and characterization of fluctuations in financial time series are subjects of considerable interest \cite{hes2,bouchad1} (and references therein). These non-stationary time series are well known to exhibit multiple characteristics depending on the scale of observation \cite{hes5}. Apart from bursty behavior, fluctuations exhibit self-similar features, with non Gaussian characteristics \cite{gs1} and inverse cubic law behavior \cite{hes2,hes5,hes6,hes7}. The advent of wavelet transform \cite{daub1} enabled one to isolate fluctuations at different scales, for their characterization. The study of fluctuation functions, both by multifractal detrended fluctuation analysis (MFDFA) and wavelet analysis reveals multi-fractal behavior \cite{hes6,hes7,mani1,mani2,mani3,mani4}. 
\par
In particular, the wavelets belonging to the Daubechies family enabled one to isolate local polynomial trends of different orders, in a manner where the fluctuations at different scales are completely independent. This is because of the exact orthogonality property of the discrete wavelet basis sets at different scales \cite{farge1}. This study has revealed non-statistical features in certain financial time series, in the context of BSE-100 price indices \cite{mani4}. The presence of correlations in various time series that can removed through shuffling is called the non-statistical property of a time series \cite{mani4}. These non statistical correlations are the artifacts of external impulses provided by the socio-political factors. The fact that, determination of multi-fractal behavior requires isolation of local fluctuations, both at low and high scales, motivates one to study their statistical characteristics more carefully. Apart from the study of their non-Gaussian properties at different scales, this analysis will also help in throwing more light on the non-statistical nature of the fluctuations. We make use of Db4 and Db6 discrete wavelets to respectively extract and study the fluctuations at multiple scales several large-cap equities listed in the New York stock Exchange, belonging to different sectors of the economy. We have systematically checked non-Gaussianity and self-similar characteristics of the multi-scale variations, while checking emergence of  $k^{-3}$ behavior of the averaged fluctuations. The study reveals the difference in the self-similar behavior of the various companies and presence of non-statistical behavior at small and large scales in certain cases. Fluctuation analysis indicates the presence of unstable, short periodic modulations. One also observes local periodic behavior of longer duration, ably identified by the Morlet wavelet. It is found that companies belonging to different sectors of the economy show distinct phase lags, in their periodic modulations. To summarize the main results of the work, in this article we show the convergence of the $\alpha \rightarrow-3$ as a function of the scale (corresponding to coarser grained analysis) and the presence of periodic phase shifted modulations of the various companies as functions of different scales. We must point out that this kind of analysis on the fluctuations to ascertain the behavior of the power law exponent and the statistical properties of the fluctuations has not been carried out before to the best of our knowledge.
\par

\section{Theory and Methodology}\label{sec:booc}
\subsection{Data}\label{sec:data}
We use the closing prices of fifteen large-cap companies listed in the New York Stock Exchange (NYSE) \cite{yahoo1} in our analysis. In this work, we limit ourselves to the behavior of the large-cap equities and are interested in the connections this specific section of the economy shows. The list of the equities is given in Table \ref{tab:companies}.
\begin{table}[h]
\begin{tabular}{l c c r}
\hline
\hline
Company & NYSE ticker & Start date & End Date \\
& symbol & MM-DD-YYYY & MM-DD-YYYY\\
\hline
Bank of America & \textbf{BAC} & 05-29-1986 & 10-02-2009\\
Boeing Co. & \textbf{BA} & 01-02-1962 & 10-02-2009 \\
General Electric Co. & \textbf{GE} & 01-02-1962 & 10-02-2009 \\
British Petroleum plc. & \textbf{BP} & 01-03-1977 & 10-02-2009 \\
Coca Cola Co. & \textbf{KO} & 01-02-1962 & 10-02-2009 \\
Pepsi Co. & \textbf{PEP} & 01-03-1977 & 10-02-2009\\
Lockheed Martin Corp. & \textbf{LMT} & 01-03-1977 & 10-02-2009\\
Walt Disney Co. & \textbf{DIS} & 01-02-1962 & 10-02-2009\\
Wal-Mart Stores Inc. & \textbf{WMT} & 08-25-1972 & 10-02-2009\\
International Business Machines Corp. & \textbf{IBM} & 01-02-1962 & 10-02-2009\\
McDonalds Corp. & \textbf{MCD} & 01-02-1970 & 10-02-2009\\
Ford Motor Corp. & \textbf{F} & 01-03-1977 & 10-02-2009\\
Alcoa Inc. & \textbf{AA} & 01-02-1962 & 10-02-2009\\
Merck and Co. Inc. & \textbf{MRK} & 01-02-1970 & 10-02-2009\\
American Express Co. & \textbf{AXP} & 04-01-1977 & 10-02-2009\\
\hline
\end{tabular}
\caption{\label{tab:companies} Summary of the data used in the analysis. The data has been obtained from \cite{yahoo1}.}
\end{table}
\subsection{Wavelet transforms}
In this work, we will use both continuous and discrete wavelet transforms for the analysis of the equities. The Continuous Wavelet Transform (CWT) is useful in analyzing the periodic modulations present in the time series at different scales. It is noteworthy to point out here that since the continuous wavelet basis is over complete, the wavelet coefficients obtained from CWT cannot be used to reconstruct the time series and hence cannot be used to extract fluctuations. We employ Discrete Wavelet Transforms (DWT) to achieve the extraction of fluctuations.
\subsubsection{Discrete Wavelet Transform (DWT)}\label{dwt_o}
Discrete wavelet transform (DWT) \cite{torrence1,farge1}, is composed of two kernels called the father wavelet $\phi(t)$ and the mother wavelet $\psi(t)$, which satisfy the following admissibility conditions \cite{daub1,mallat1}:
\begin{eqnarray}
\int \phi dt < \infty , \int \psi dt &=& 0 , \int \phi^{*} \psi dt =0,\\
\int \vert \phi \vert^2 dt =\int \vert \psi \vert^2 dt &=& 1,
\end{eqnarray}
The daughter wavelets (which are scaled and translated versions of the mother wavelet)
\begin{equation}
\psi_{j,k}(t)=2^{j/2} \psi(2^j t-k), \qquad k \in \mathbb{R},\quad j \in \mathbb{Z}^+
\end{equation}
differ from the mother wavelet in terms of the height and the width which, at the $j^{th}$ scale differ by a factor of $2^j$ and $2^{j/2}$ respectively. $j$ and $k$ are the scaling parameters and are real and positive integers respectively. 
\par
The discrete wavelet decomposition of a function $f(t)$ is given by,
\begin{equation}
f(t)=\sum_{k=-\infty}^{\infty} c_{k} \phi_{k}(t) +\sum_{k=-\infty}^{\infty} \sum_{j \geq 0} d_{j,k} \psi_{j,k} (t)
\end{equation}
where $c_{k}$s are the low pass coefficients, which capture the trend (average behavior or the long wavelength components) and $d_{j,k}$s are the high pass coefficients that extract the high frequency components (or fluctuations) of the function. Multi Resolution Analysis (MRA) \cite{daub1} leads to,
\begin{eqnarray}
c_{j,k}=\sum_{n} h(n-2k) c_{j+1,n}\\
d_{j,k} =\sum_{n} \tilde{h}(n-2k)c_{j+1,n}
\end{eqnarray}
where, $h(n)$ and $\tilde{h}(n)$ are the low pass and high pass filter coefficients respectively. It is worth emphasizing that low pass and high pass coefficients at a given scale can be obtained by low pass coefficients alone at a higher scale. Thus, we can go from a high resolution analysis of the function to a progressively coarse grained picture by convolution with the filter coefficients.

\subsubsection{Continuous Wavelet Transform (CWT)}\label{cwt_o}
Continuous wavelet transform was pioneered by Morlet \cite{farge1}. An integrable, well localized (in both the physical and Fourier domains), zero mean function, the mother wavelet $\psi(n)$, is used as the analyzing function. Given a discrete data set $X = \{x_n\colon n\in \mathbb{Z}^+\}$, the wavelet coefficients are calculated by convolving the $\dat$, with the scaled and translated $\psi(n)$. The wavelet coefficients are given by,
\begin{equation}
W_n(s)=\sum_{n'=0}^{N-1} x_{n'} \psi^{*}\left(\frac{n-n'}{s} \right)
\end{equation}
where $s$ is the scale. The admissibility conditions are given by,
\begin{eqnarray}
\label{eq:eq2}
\int_{\mathbb{R}^r} \vert\hat{\psi}(\vec{k})\vert^2 \frac{d^r \vec{k}}{\vert \vec{k} \vert^r} &<& \infty, \\
\label{eq:eq3}
\mbox{where}, \hat{\psi}(\vec{k}) &=& \frac{1}{(2\pi)^r} \int_{\mathbb{R}^r} \psi(\vec{x})e^{-\imath \vec{k}\cdot\vec{x}}d^r \vec{x}\\
\label{eq:eq4}
\mbox{and }\int_{\mathbb{R}^r}\psi(\vec{x})d^r \vec{x}&=&0.
\end{eqnarray}
Here $r$ is the number of spatial dimensions.
In our analysis, we use the complex Morlet wavelet given by,
\begin{equation}
\label{eq:eq5}
\psi(n)=\pi^{-1/4}e^{\imath \omega_0 n}e^{-n^2 /2}
\end{equation}
where, n is the localized time index. Though Eq.(\ref{eq:eq5}) is a marginally admissible function, it is made admissible by taking $\omega_0=6$. The Fourier wavelength of $\psi(n)$, $\lambda$, is given by,
\begin{equation}
\label{eq:lamb_morl}
\lambda=\frac{4\pi s}{\omega_0+\sqrt{2+\omega_0^2}} \sim 1.03s
\end{equation}
implying that the scale and the wavelength for the Morlet wavelet are approximately equal. The scale and frequency $\nu$ are related by,
\begin{equation}
\nu \propto \frac{1}{s}.
\end{equation}
The cone of influence (COI) defined in \cite{stephane1999wavelet}, is the region, beyond which, convolution errors make the wavelet coefficients unreliable for analysis \cite{torrence1}. 
\begin{figure}
\centering
\subfigure[CWT of \textbf{DIS} from scale 384 to 768]{
\includegraphics[scale=0.30]{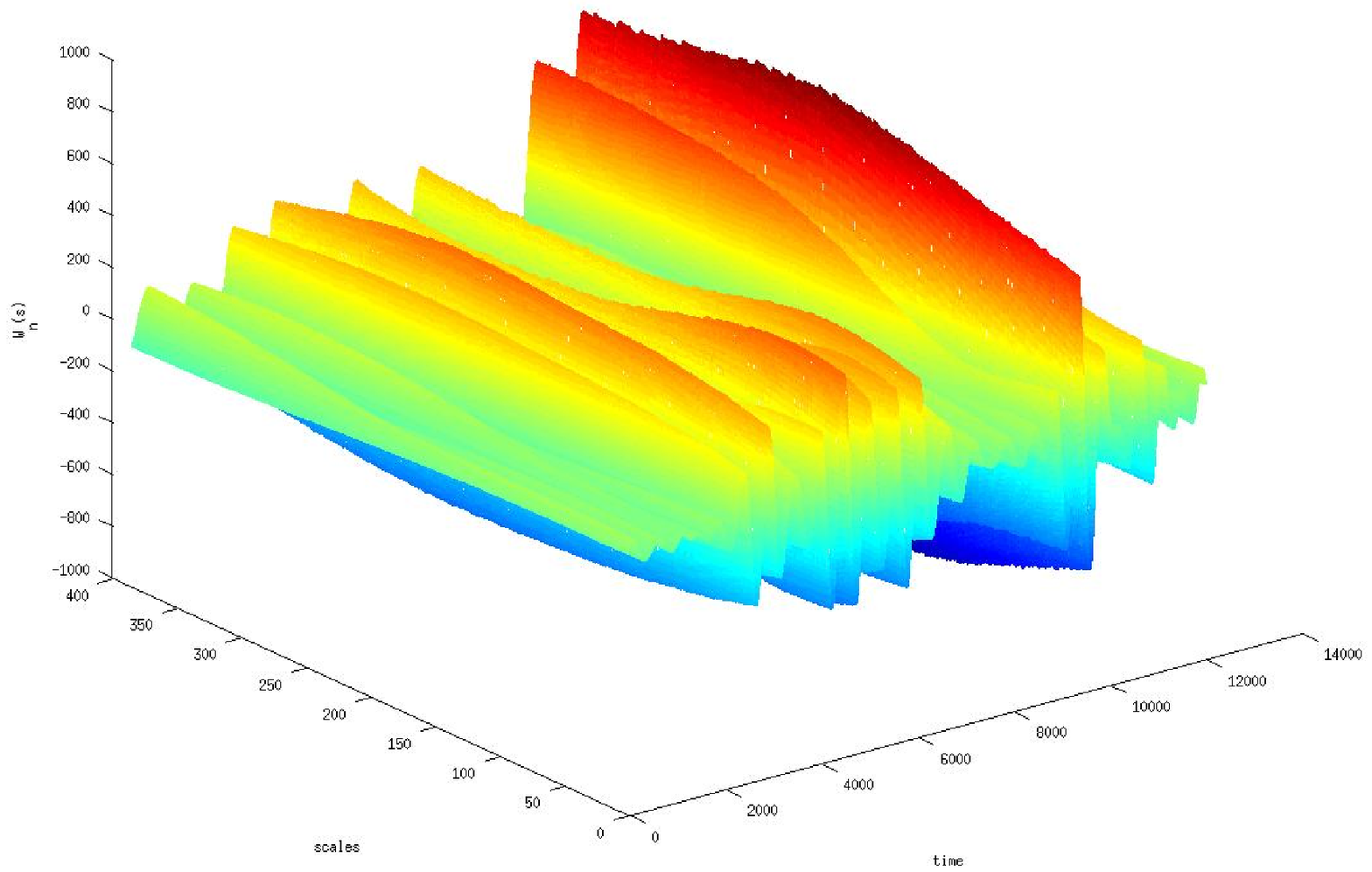}
\label{fig:dis_cwt}
}
\subfigure[CWT of \textbf{KO} from scale 384 to 768]{
\includegraphics[scale=0.30]{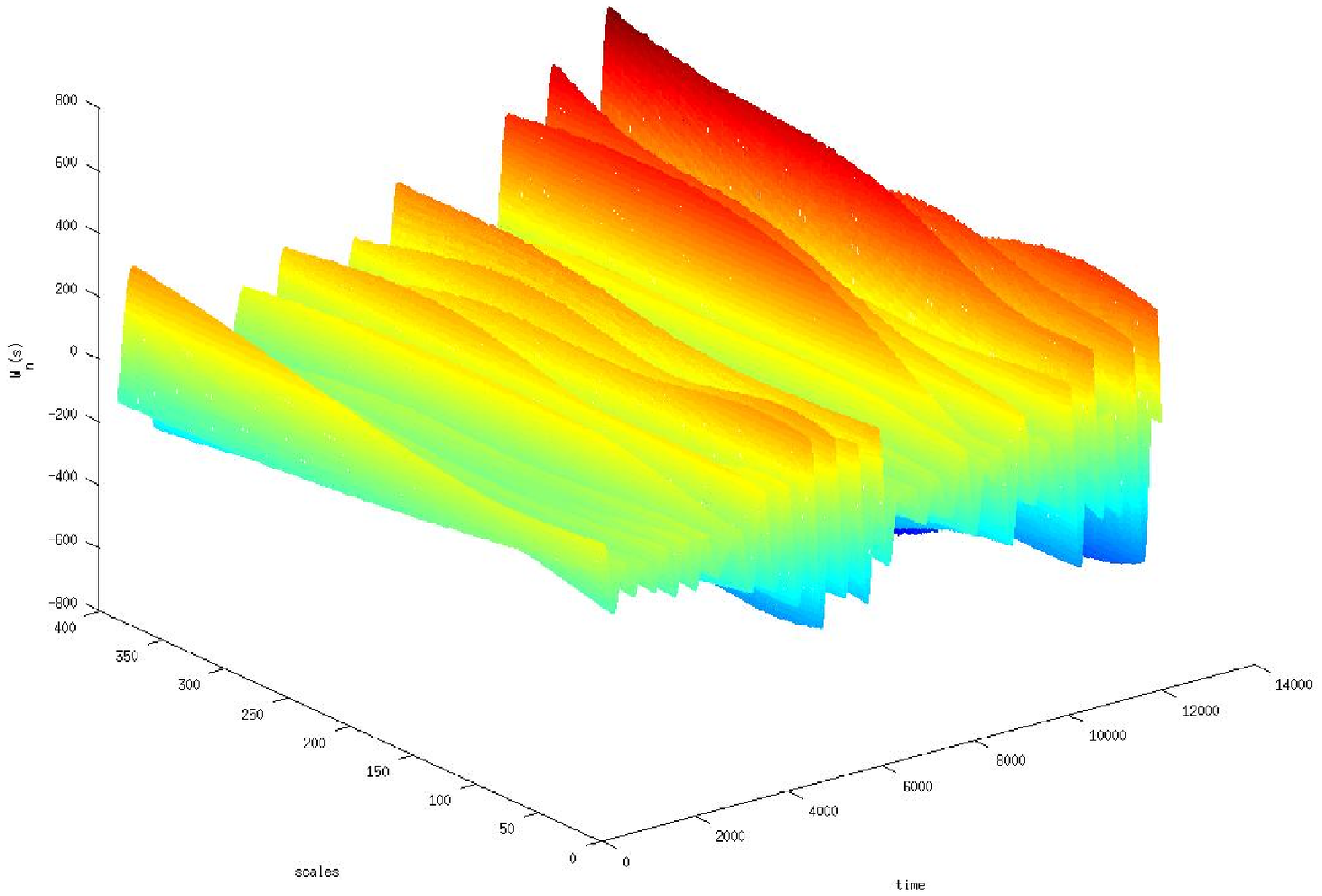}
\label{fig:ko_cwt}
}
\caption{\label{fig:cwt_2_comp}The scalograms representing the coefficients of the Morlet wavelets of the companies \subref{fig:dis_cwt} \textbf{DIS} and \subref{fig:ko_cwt}\textbf{KO}, from scales 384 to 768, the x-axis represents time, y-axis scales and z-axis wavelet coefficients $W_n (s)$.} Periodic modulations are clearly visible.
\end{figure}

\subsection{Returns and autocorrelation function}\label{sec:acf_ret_sec}
It is well known that for a given time series $\{x_n\}$, a slowly decaying autocorrelation function given by,
\begin{equation}
S(n)=\frac{1}{N-n-1} \sum_{t=1}^{N-n} \frac{(x_n-\mu_1)(x_{n+\tau}-\mu_2)}{\sigma_1 \sigma_2}
\end{equation}
signifies non-stationarity in $\{x_n\}$ \cite{bragg1,burke1}. Here, $S(n)$ is the correlation for the lag $n \in [0, 3N/4]$, where N is the length $\{x_n\}$ and $3N/4$ is the maximum allowed lag \cite{burke1}. $\mu_1,\mu_2$ are the mean of $\{x_1,\ldots,x_{N-n}\}$ and $\{x_{n+1},\ldots,x_N\}$ respectively and $\sigma_1,\sigma_2$ are the corresponding standard deviations.
\begin{figure}
\centering
\subfigure[Autocorrelation function (ACF) of \textbf{AA}]{
\includegraphics[scale=0.35]{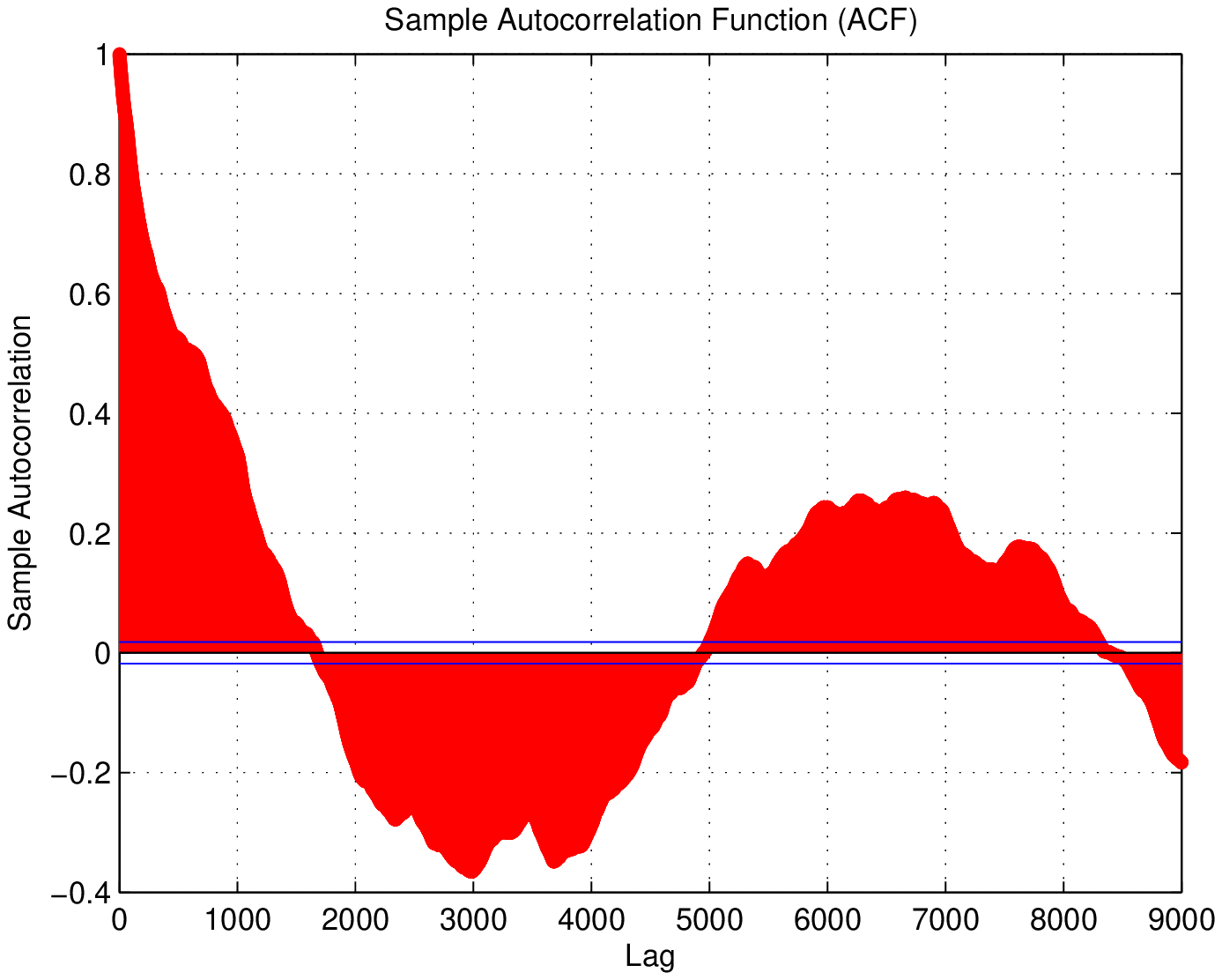}
\label{fig:acf_alcoa}
}
\subfigure[ACF of the normalized log returns of \textbf{AA} showing zero correlation]{
\includegraphics[scale=0.35]{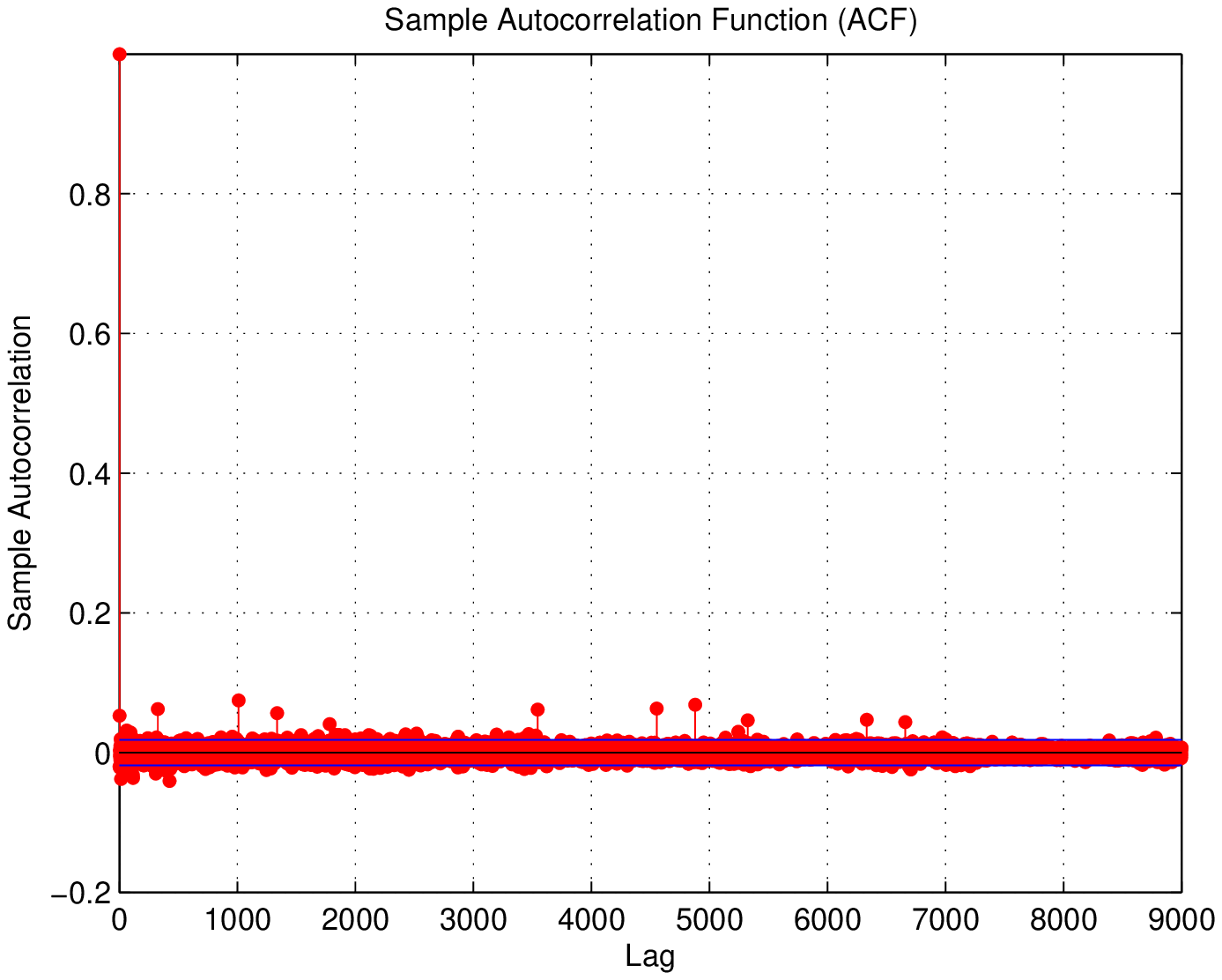}
\label{fig:acf_alcoa_norm_ret} 
}
\caption{\label{fig:acf} In \subref{fig:acf_alcoa} ones observes the non stationary behavior of a typical time series, while in \subref{fig:acf_alcoa_norm_ret}, we find that the normalized log returns are stationary.}
\end{figure}

However, a differencing operation, like log normalized returns can make the corresponding time series stationary \cite{burke1}, i.e., the autocorrelation function reaches the level of noise very quickly. The normalized log returns (NLR), $\hat{R}$ are defined as,
\begin{equation}
\label{eq:norm_ret}
\hat{R}=\frac{R(n)-\langle R(n) \rangle}{\sqrt{\langle R(n)^2 \rangle - \langle R(n) \rangle^2}}
\end{equation}
where, $R(n)=\log x_{n+1}-\log x_n$ and $\sqrt{\langle R(n)^2 \rangle - \langle R(n) \rangle^2}$ is called the volatility of returns. We consider the NLR for further analysis. We observe that though the stock prices show non-vanishing autocorrelation (Fig.\ref{fig:acf_alcoa}), the auto correlation function of the normalized log returns tends to zero very quickly, signifying stationarity(Fig.\ref{fig:acf_alcoa_norm_ret}). This behavior is observed in the time series of all the companies. We concentrate on the self similar characteristics of the fluctuations and use the Daubechies' family of wavelets to this end.
\subsection{Wavelet based fluctuation extraction and analysis and characterization of multifractality} \label{sec:fluc_extr}
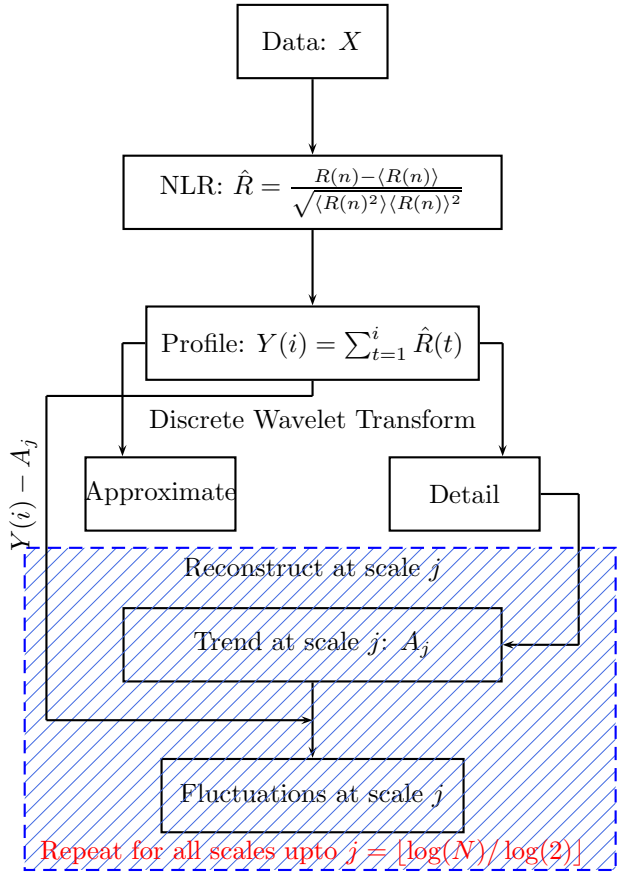
\begin{figure}
\begin{center}
\psset{unit=1cm}
\begin{pspicture}(-2.5,-2.5)(9,9)
\newrgbcolor{royalblue}{0.25 0.41 0.88}
\psframe(4,8)(6,9)
\rput(5.0,8.5){Data: $\dat$}
\psframe(2.5,6)(7.5,7)
\psline{->}(5,8)(5,7)
\psline{->}(5,6)(5,5)
\rput(5.0,6.5){NLR: $\hat{R}=\frac{R(n)-\langle R(n)\rangle}{\sqrt{\langle R(n)^2\rangle \langle R(n)\rangle^2}}$}
\psframe(2.8,4)(7.2,5)
\rput(5.0,4.5){Profile: $Y(i)=\sum_{t=1}^{i}\hat{R}(t)$}
\rput(5.0,3.5){Discrete Wavelet Transform}
\psframe(2,2)(4,3)
\psframe(6,2)(8,3)
\rput(3,2.5){Approximate}
\rput(7,2.5){Detail}
\psline{-}(2.8,4.5)(2.5,4.5)
\psline{->}(2.5,4.5)(2.5,3.0)
\psline{-}(7.2,4.5)(7.5,4.5)
\psline{->}(7.5,4.5)(7.5,3.0)
\psframe(2.5,0)(7.5,1)
\rput(5.0,1.5){Reconstruct at scale $j$}
\psline{-}(8.5,2.5)(8.5,0.5)
\psline{-}(8,2.5)(8.5,2.5)
\psline{->}(8.5,0.5)(7.5,0.5)
\rput(5.0,0.5){Trend at scale $j$: $A_j$}
\psline{->}(5,0)(5,-1)
\psframe(3,-2)(7,-1)
\rput(5.0,-1.5){Fluctuations at scale $j$}
\psframe[linecolor=blue,linestyle=dashed,fillstyle=hlines,hatchwidth=0.01cm,hatchcolor=royalblue](1.2,-2.5)(9.0,1.8)
\rput(5.0,-2.25){\red Repeat for all scales upto $j=\lfloor \log(N)/\log(2) \rfloor$}
\psline{-}(5,3.8)(5,4)
\psline{-}(5,3.8)(1.5,3.8)
\psline{-}(1.5,3.8)(1.5,-0.5)
\psline{->}(1.5,-0.5)(5,-0.5)
\rput{90}(1.2,2.5){$Y(i) - A_j$}
\end{pspicture}
\end{center}
\caption{\label{fig:algo} This figure shows the algorithm for extracting fluctuations from the data.}
\end{figure}

The Daubechies family of wavelets are made to satisfy vanishing moments conditions,
\begin{equation}
\int  t^n \psi_{j,k} (t)  dt=0, \quad (n=1,\ldots,N)
\end{equation}
due to which this family of wavelets extracts polynomial trends. For example, with Db4, where, $n=1$, the wavelet coefficients are blind to a linear trend. These are captured by the low-pass coefficients. A discussion on this can be found in \cite{daub1}. We use this property to extract the fluctuations from $\dat$ by reconstructing the original signal at each level using the approximation coefficients and then taking the difference of the reconstructed signal from the original signal. The algorithm used is shown in Fig.(\ref{fig:algo}).
\begin{figure}
\centering
\includegraphics[scale=0.50]{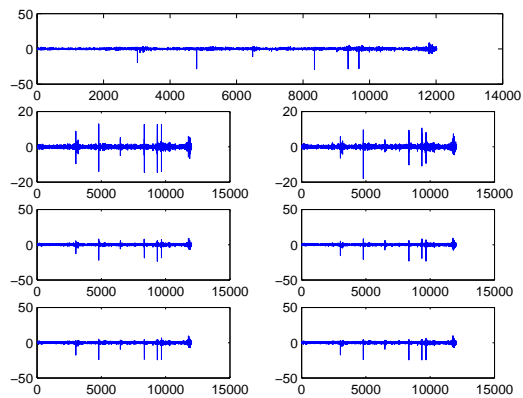}
\caption{\label{fig:fluctuations_fig} The top plot depicts the normalized log returns of the closing stock price \textbf{AA}. The left column shows the fluctuations at levels 1, 2 and 3 respectively, using Db4 wavelet. The right column shows the corresponding results obtained by using Db6 wavelet.}
\end{figure}
Fig.(\ref{fig:fluctuations_fig}) depicts the NLR and the fluctuations extracted by the wavelet based method. The top plot shows the NLR, the left and right columns are obtained using Db4 and Db6 wavelets respectively. We observe the transient and bursty behavior at different levels of analysis which were not apparent from the NLR.
\par
The discrete wavelet based method to characterize multifractality proposed by Manimaran \textit{et al.} \cite{mani1,mani2}, has been successfully applied to extract multifractality of various time series \cite{mani3,mani4,sayantan1}. In this procedure, we use the normalized log returns obtained through Eq(\ref{eq:norm_ret}) to obtain the time series profile by taking a cumulative sum of the normalized returns:
\begin{equation}
Y(i)=\sum_{t=1}^{i} \hat{R}(t),\quad i=1,\ldots,N-1.
\end{equation}
Then, a DWT is applied on the profile, $Y(i)$, to obtain the fluctuations of the profile at different scales. We use Db4 and Db6 to extract polynomial trends which are respectively locally linear and quadratic. The edge effects and the asymmetric nature affect the precision of the extracted fluctuations and hence, we use a DWT on the reverse profile to extract a new set of fluctuations. This new set of fluctuations are then reversed and averaged over the earlier obtained fluctuations to get the set of fluctuations for subsequent analysis. 
\par
We subdivide the fluctuations into non-overlapping segments $M_s=\mbox{int}(N/s)$, $N$ and $s$ being the length of the fluctuations and the scale respectively. The $q^{th}$ order fluctuation function $F_q(s)$ is then obtained by squaring and averaging the fluctuations over all the segments:
\begin{equation}
F_q(s)\equiv \left[ \frac{1}{2M_s}\sum_{j=1}^{2M_s}\left( F^2(j,s)\right)^{\frac{q}{2}} \right]^{\frac{1}{q}},
\end{equation}
where, $q$ is the order of the moment. We repeat the above procedure for different scales $s$ for different values of $q$ except $q=0$. We thus obtain the power law scaling behavior from the fluctuation function as,
\begin{equation}
F_q (s) \sim s^{h(q)}
\end{equation}
in a logarithmic scale for each value of $q$. At $q=0$, due to the divergence of the scaling exponent, a logarithmic averaging is employed to find the fluctuation function;
\begin{equation}
F_q(s) \equiv \exp \left[ \frac{1}{2M_s}\sum_{j=1}^{2M_s}\log \left( F^2(j,s)\right)^{\frac{q}{2}} \right]^{\frac{1}{q}}.
\end{equation}
For mono fractals time series \cite{mandelbrot1,mandelbrot2}, $h$ is independent of $q$, while for multi fractals, $h \equiv h(q)$. The Hurst scaling exponent, $H$, a measure of the fractal nature of the time series is calculated as $h(q=2)=H$ and $H \in [0,1]$, where, $H \in [0,0.5)$ and $H \in (0.5,1]$ reveal, respectively, the anti-persistent and persistent nature  of the time series.
\subsection{Tests for Gaussianity}\label{sec:gaussianity}
For a random variable $x$, the cumulative distribution function (CDF) is given by,
\begin{equation}
F(x)=\int_{-\infty}^{x} f(n) dn
\end{equation}
where, $f(n)$ is the probability distribution function of $\dat$. The Kolmogorov-Smirnov test (KS-test) \cite{kolmogorov1,smirnov1} is used to compare $F(x)$ of an empirical time series with that of a normal or Gaussian distribution $F'(x)$. The KS-statistic is given by,
\begin{equation}
D_n=\sup_n \vert F'(x)-F(x) \vert
\end{equation}
where, $\sup$ is the supremum defined as the lowest element of a subset $S$ of a partially ordered set $T$ which is greater than all other elements of $S$. For a continuous $F(x)$, under the null hypothesis that it comes from $F'(x)$, the quantity
\begin{equation}
\label{eq:kstheorem}
\sqrt{n}D_n \xrightarrow{n\rightarrow \infty} \sup_n \vert B(F(x)) \vert
\end{equation}
converges to the Kolmogorov distribution given by, 
\begin{equation}
K=\sup_{n \in[0,1]} B(n)
\end{equation}
 where, $B(n)$ is the Brownian bridge or a Brownian process. The Kolmogorov distribution is independent of $F(x)$\cite{melo1,eadie1,massey1,miller1,marsaglia1,campbell1}. In the present case, the KS-test indicated non-Gaussian nature for fluctuations obtained through wavelet transforms.
\par
\begin{figure}
\centering
\subfigure[db4]{
\includegraphics[scale=0.20]{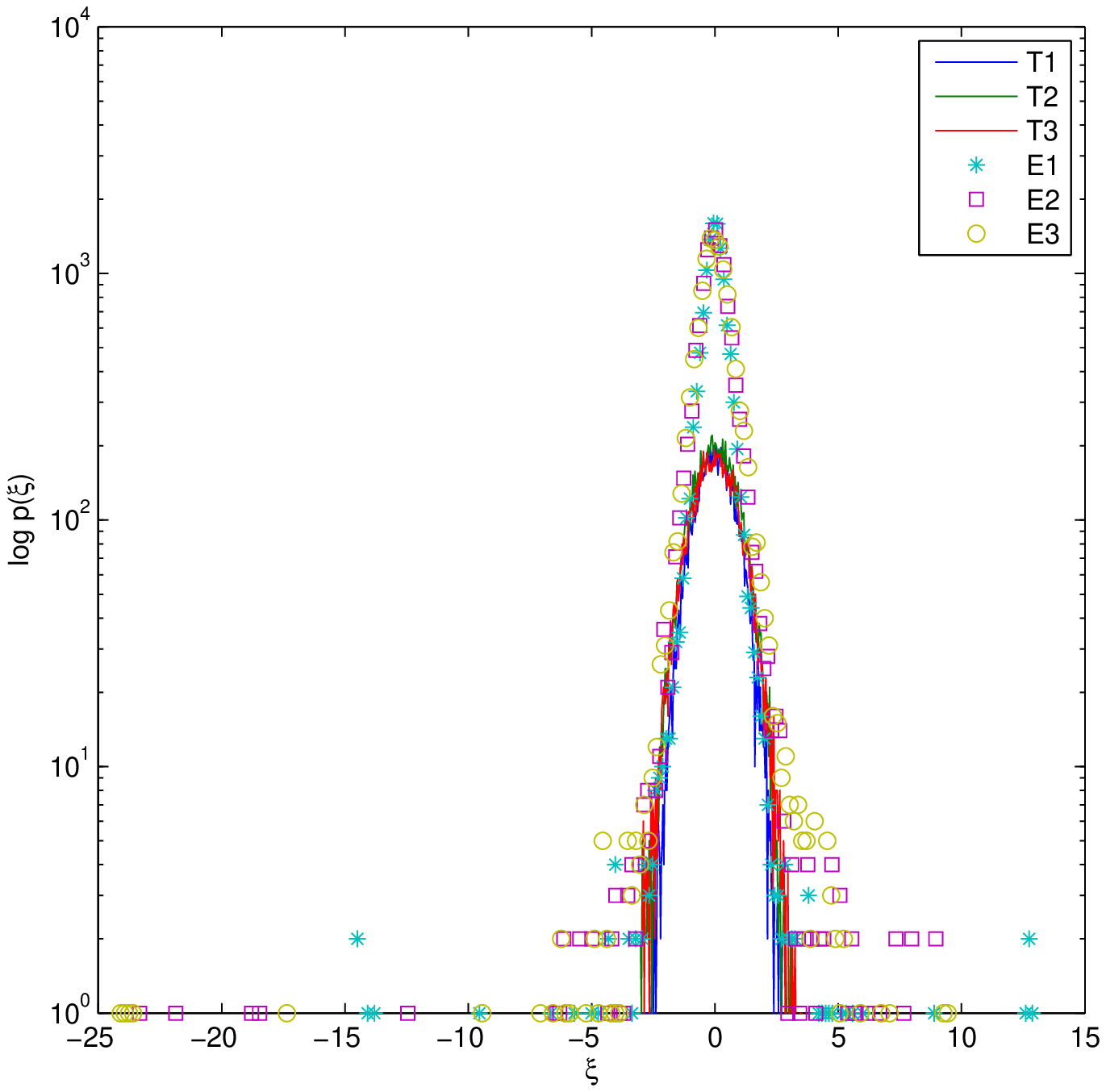}
\label{fig:hf41}
}
\subfigure[db6]{
\includegraphics[scale=0.20]{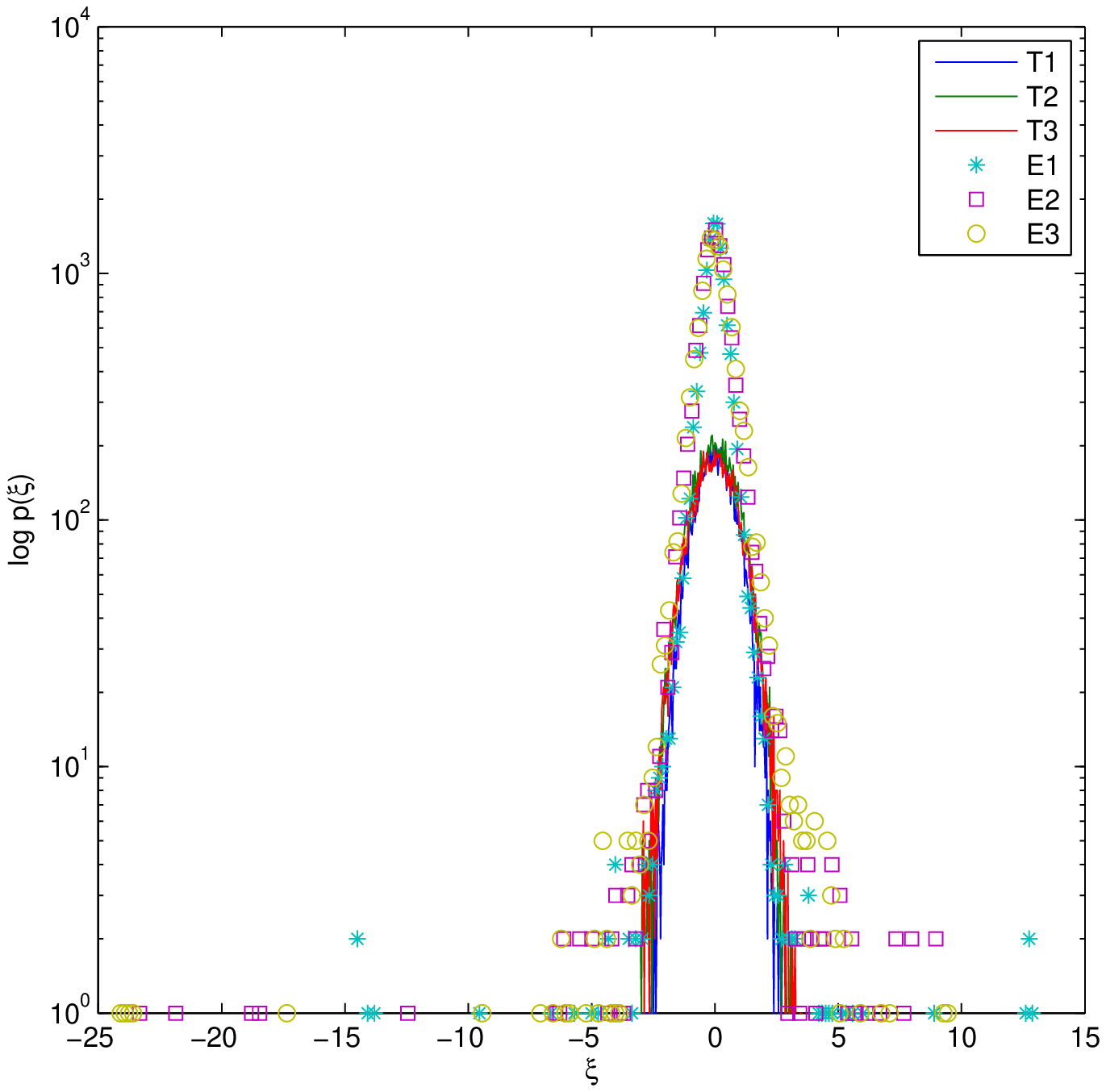}
\label{fig:hf42}
}
\subfigure[Skewness ($\gamma$)]{
\includegraphics[scale=0.35]{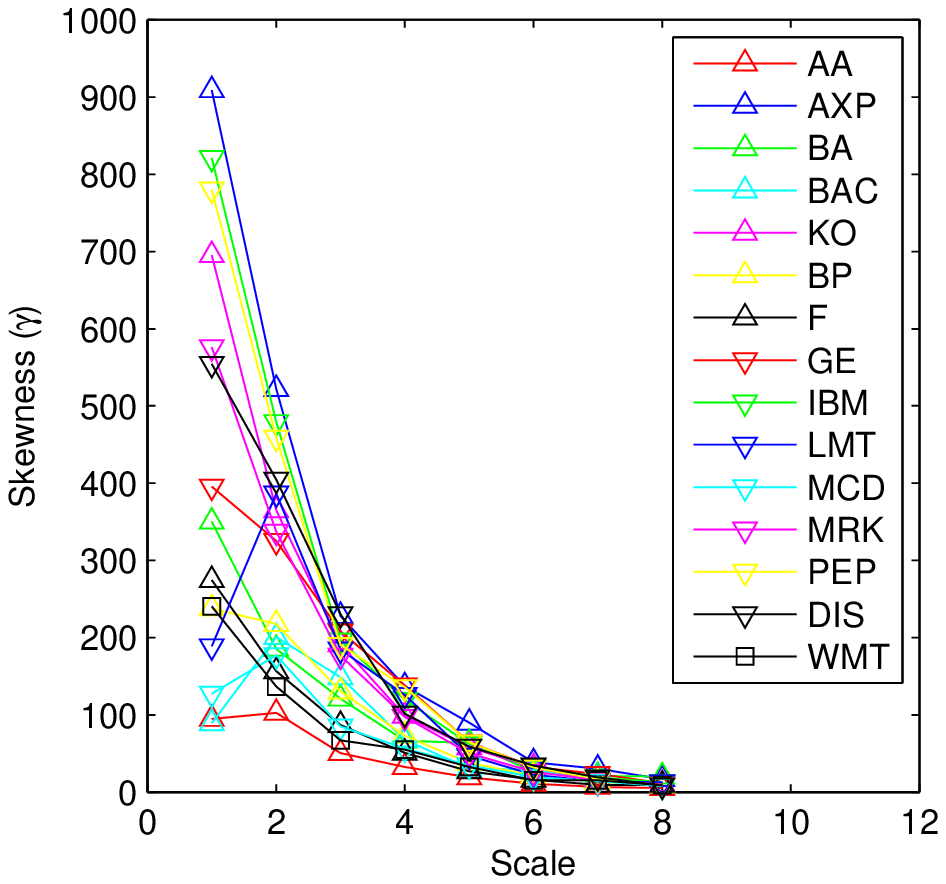}
\label{fig:skew_all}
}
\subfigure[Kurtosis ($\kappa$)]{
\includegraphics[scale=0.30]{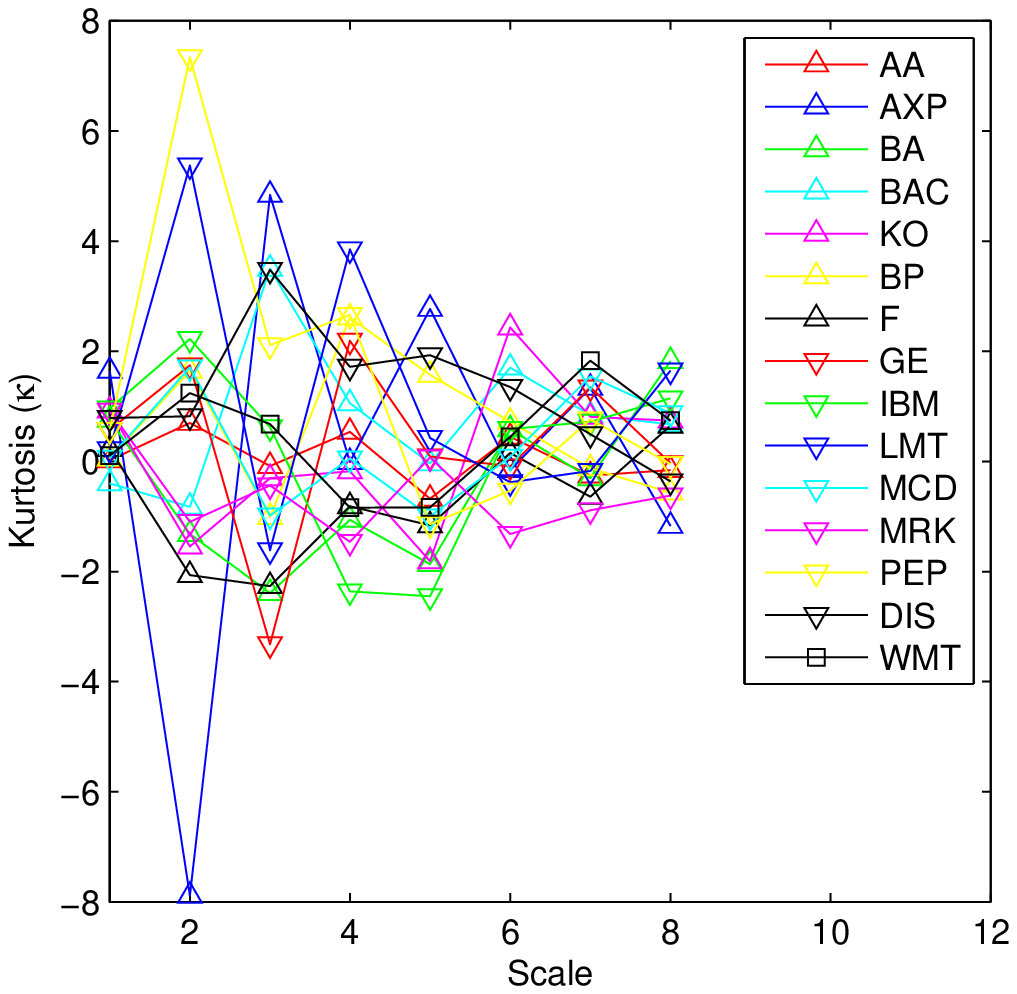}
\label{fig:kurt_all}
}

\caption{\label{fig:gauss} \subref{fig:hf41} and \subref{fig:hf42} depict the distribution of the fluctuations at each level against a normal distribution taken with the same mean and standard deviation as the fluctuations. T1, T2 and T3 are the theoretical distributions and E1, E2, E3 are the fluctuations at levels 1, 2 and 3 respectively. \subref{fig:skew_all} and \subref{fig:kurt_all} show skewness ($\gamma$) and kurtosis ($\kappa$) against scale respectively.}
\end{figure}

However, we observe that though the fluctuations show non-Gaussian behavior(see Figs.\ref{fig:hf41},\ref{fig:hf42}), sufficient averaging at the higher scales tends to bring the fluctuations towards a Gaussian behavior (skewness$=\gamma=0$ and kurtosis$=\kappa=3$), as shown in Figs.\ref{fig:skew_all} and \ref{fig:kurt_all}. 
\subsection{Power law behavior of the fluctuations}\label{sec:powerlaw}
A random variable $x$ follows a power law if it belongs to a family of distributions of the form 
\begin{equation}
p(x) \propto L(x)x^{\alpha},
\end{equation}
where, L(x) is a slowly varying function such that 
\begin{equation}
\lim_{x \rightarrow \infty} \frac{L(cx)}{L(x)} =1
\end{equation}
for $c>0$. The It is observed (see Fig. \ref{fig:power_law}) that the $k^{-3}$ behavior emerges at higher scales. Also, it can be seen from Fig. \ref{fig:power_law}, that at the scale $5$, all the companies seem to converge to the same value of $\alpha$ and diverge at higher scales. 

\section{Multi-fractality and non-statistical behavior}
\begin{figure}
\centering
\subfigure[Fourier Power of IBM versus scales.]{
\includegraphics[scale=0.25]{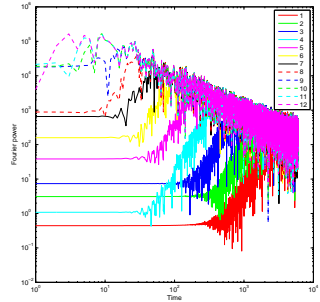}
\label{fig:fourier_power}
}
\subfigure[Power law exponent $\alpha$ versus scale.]{
\includegraphics[scale=0.45]{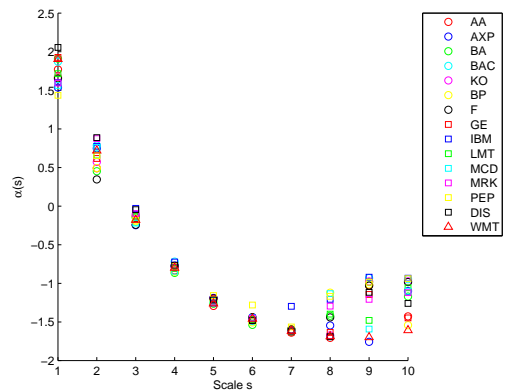}
\label{fig:power_law}
}
\caption{\label{fig:power} \subref{fig:fourier_power} shows the Fourier power spectrum of \textbf{IBM} for various scales and \subref{fig:power_law} depicts the power law exponent ($\alpha$) versus scale. The high frequency components reveal $1/k^3$ behavior. The rising component in the medium frequency range indicates unstable periodic modulations.}
\end{figure}

\begin{figure}
\centering
\subfigure[Unshuffled returns (Db4)]{
\includegraphics[scale=0.30]{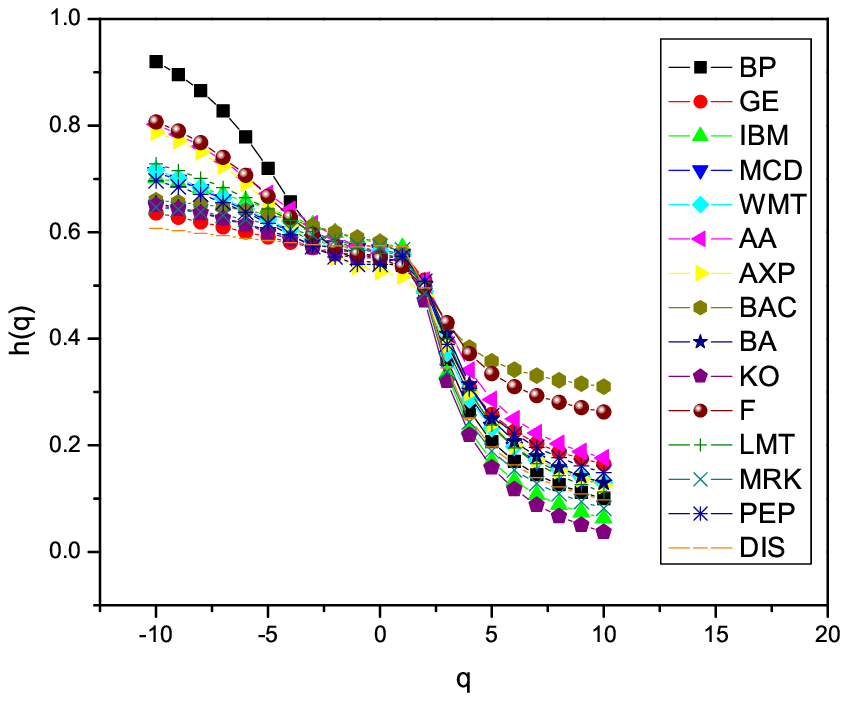}
\label{fig:hurst_a}
}
\subfigure[Unshuffled returns (Db6)]{
\includegraphics[scale=0.30]{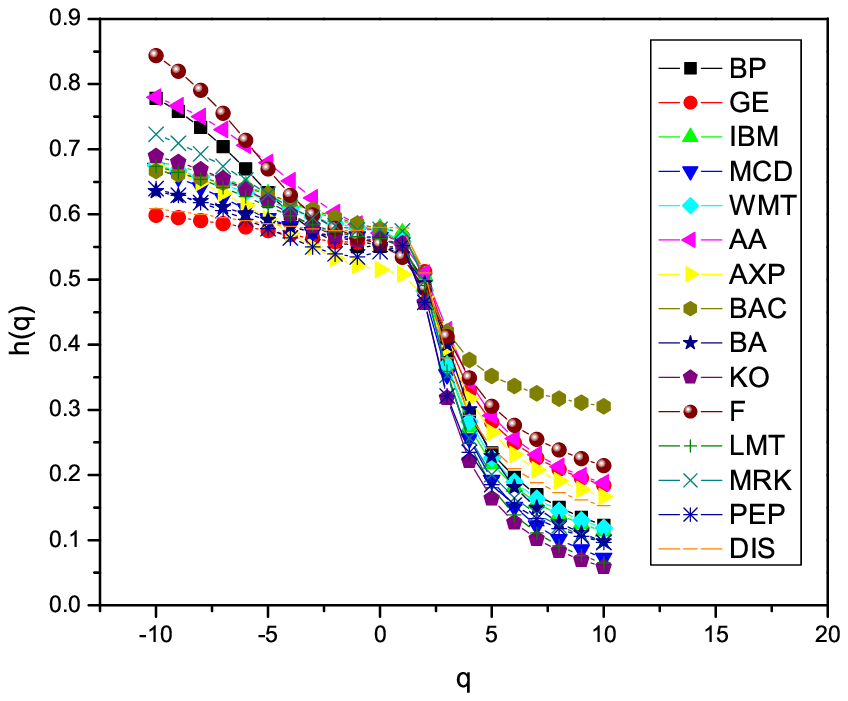}
}
\subfigure[Shuffled returns (Db4)]{
\includegraphics[scale=0.30]{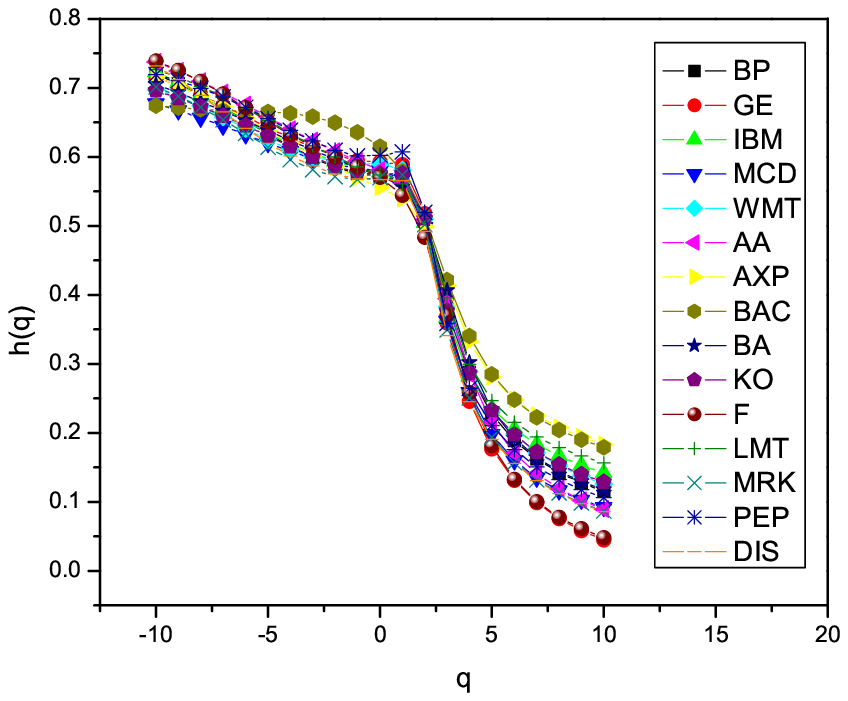}
}
\subfigure[Shuffled returns (Db6)]{
\includegraphics[scale=0.30]{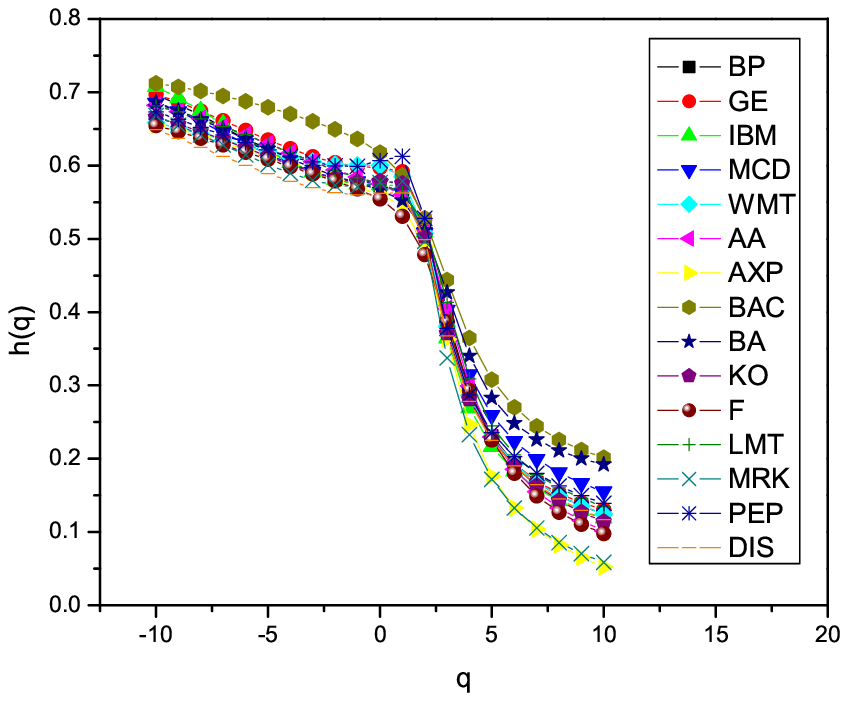}
}
\caption{\label{fig:hurst}$h(q)$ vs $q$ for unshuffled and shuffled returns.}
\end{figure}

As described earlier, $Db4$ and $Db6$ wavelets have been used here to extract fluctuations at multiple scales, from which the corresponding fluctuation functions $F_q(s)$ are constructed. The $h(q)$ values are extracted from the $F_q(s)\sim s^{h(q)}$, which have been plotted in Fig.\ref{fig:hurst} as a function of $q$ ranging from $-10$ to $10$. It is worth mentioning that the smaller values of $q$ get dominant contributions from small functions, whereas the large fluctuations contribute significantly for larger $q$ values. Multi-fractality is clearly observed in all the return time series. Interestingly, the $h(q)$-values extracted from the shuffled time series, showed different behavior for some companies, particularly at low and high $q$-values. This reveals the non-statistical nature of the fluctuations of the corresponding companies \cite{santha1}. 
We carried out a Fourier spectral analysis to independently verify the power law behavior of the fluctuations. The cumulative of the time series revealed power-law behavior in the frequency domain, as is clearly seen in Fig.\ref{fig:power}. Very interestingly, one observes a rising component in the mid-frequency domain. This is indicative of short, unstable periodic modulations, as has been seen in a characteristic of chaotic dynamical system, as unstable periodic orbits.

The dominant low-frequency periodic components can be economically extracted through the Morlet wavelets. The scalograms reveal two dominant modulations, depicted in Fig.\ref{fig:cwt_2_comp}. One clearly observes, phase-lags between the periodic modulations of companies belonging to different sectors of the economy. One also finds that the periodic modulation at scale 512, is non-stationary, showing varying amplitudes at different times. In comparison, the periodic modulations at the scale 2048, is more uniform, being present, all through  this time domain, under investigation. The distinct phase lags physically arises from the interdependence of different companies with each other. This has also manifested in the study of the fluctuation characteristics of different companies through random matrices \cite{stanley1999,laloux1999,mitra1999}.
 From Fig.\ref{fig:power_law}, it can be observed that the $k^{-3}$ behavior emerges at higher scales. It is to be emphasised that the rising component could be related to unstable periodic orbits in the dynamical time series \cite{santha1}. 


\begin{figure}
\centering
\subfigure[Scale=512]{
\includegraphics[scale=0.20]{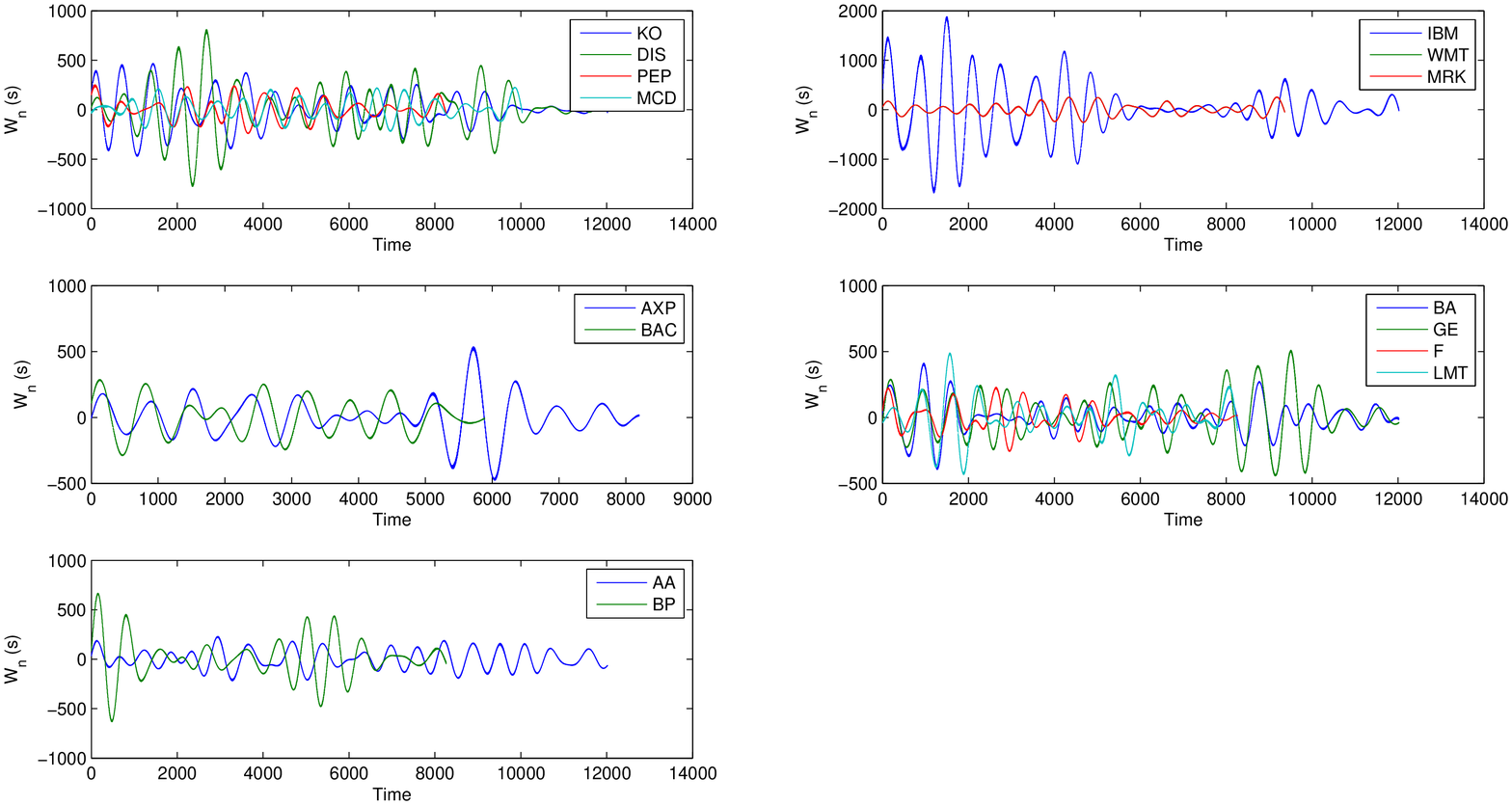}
\label{fig:cwt_plot_512}
}
\subfigure[Scale=2048]{
\includegraphics[scale=0.20]{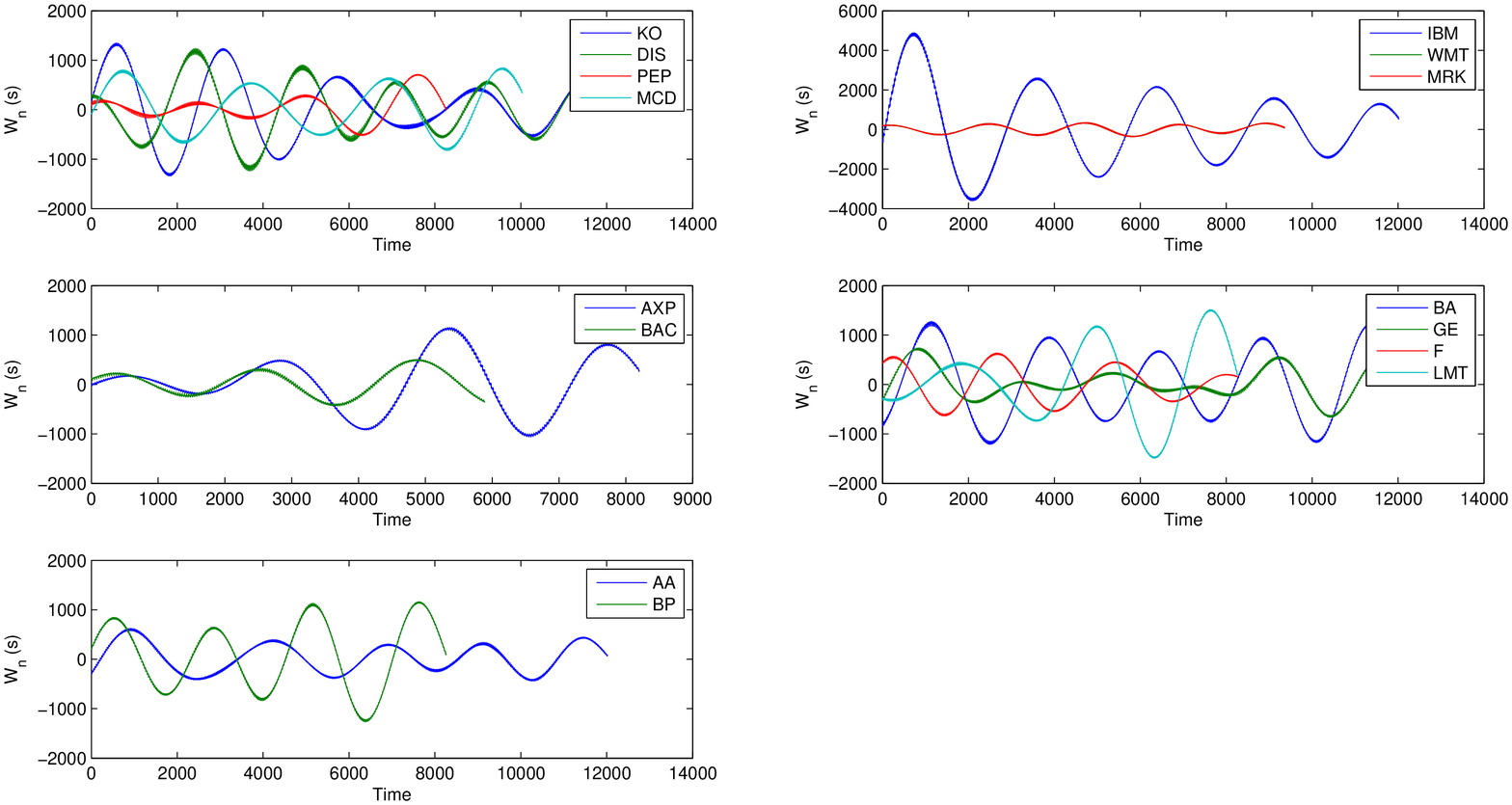}
\label{fig:cwt_plot_2048}
}
\caption{\label{fig:cwt_plot} The significant Morlet wavelet coefficients at scales \subref{fig:cwt_plot_512} 512 and \subref{fig:cwt_plot_2048} 2048. The periodic behavior of the markets at higher scales, suggesting a long range correlation is apparent. Distinct phase lags between different companies are clearly seen.}
\end{figure}

\section{Conclusion}
In conclusion, we have made use of discrete and continuous wavelets to characterize the fluctuations in the return series of a number of large companies. As expected, the time series revealed non-Gaussian behavior at high-frequencies and after sufficient averaging, Gaussian behavior manifested in certain cases. Interestingly, the Fourier spectral analysis revealed the presence of short duration unstable modulations, akin to the ones observed in chaotic dynamical systems. The multi-fractal analysis showed, non-statistical nature of the fluctuations, for certain companies, both for high and low frequency fluctuations. It is observed that, after sufficient averaging the $k^{-3}$ power law behavior emerges, when the contribution from the short, periodic modulations disappear. The Kolmogorov-Smirnov test performed to determine the Gaussianity of the fluctuations rejected the null hypothesis that the fluctuations belonged to a uniform distribution, for all the time series. It was found that the skewness and kurtosis tend towards the values for a uniform distribution ($\gamma \approx 0$ and $\kappa=3$), as the scale increases. The Morlet wavelet identified two dominant periodic modulations, where different companies, showed distinct phase lags. This suggests, a systematic study of these periodic components, by clustering, various companies into distinct groups. Random matrix approach to financial time series has also yielded, distinct groups, corresponding to different sections of the economy. Combining wavelet based approach, with the random matrix one, will lead to a better understanding of the inter-relationship of different company stock values. The nature and cause of the non-statistical components of the fluctuations, as manifested in the difference in the $h(q)$ values at small and large frequencies, between shuffled and un-shuffled time series needs further analysis. These questions are currently under investigation.

\par
\section*{Acknowledgements} SG would like to thank the hospitality of IISER Kolkata where a part of this work was done. PM would like to thank the Dept. of Science and Technology for their financial support (DST-CMS GoI Project No. SR/S4/MS:516/07 Dated 21.04.2008).
\hspace{1.5cm}
\section*{References}
\bibliographystyle{elsarticle-num}
\bibliography{myref.bib}

\end{document}